\documentclass[12pt]{article}

\usepackage{paper2e}
\usepackage{mydefs2e}

\newcommand{\LUV}{\Lambda_{\rm UV}}
\newcommand{\LIR}{\Lambda_{\rm IR}}
\newcommand{\Pt}{{\tilde{\Phi}}}
\newcommand{\Ft}{{\tilde{F}}}
\renewcommand{\d}{\partial}
\newcommand{\dd}{\raisebox{1.2pt}{$\stackrel{\raisebox{-1pt}%
{$\scriptscriptstyle\leftrightarrow$}}{\d}$}}

\newcommand{\MP}{M_{\rm P}}

\begin{document}

\begin{titlepage}
\preprint{UMD-PP-03-70}

\title{Supersymmetry without Supersymmetry}

\author{Hock-Seng Goh,%
\footnote{{\tt hsgoh@physics.umd.edu}}%
\ \ Markus A. Luty,%
\footnote{{\tt luty@umd.edu}}%
\ \ Siew-Phang Ng%
\footnote{{\tt spng@physics.umd.edu}}}

\address{Department of Physics, University of Maryland\\
College Park, Maryland 20742, USA}

\begin{abstract}
We investigate the possibility that supersymmetry is not a fundamental
symmetry of nature, but emerges as an accidental approximate
global symmetry at low energies.
This can occur if the visible sector is non-supersymmetric at
high scales, but flows toward a
strongly-coupled superconformal fixed point at low energies;
or, alternatively, 
if the visible sector is localized near the infrared brane
of a warped higher-dimensional spacetime with supersymmetry broken
only on the UV brane.
These two scenarios are related by the AdS/CFT correspondence.
In order for supersymmetry to solve the hierarchy problem, 
the conformal symmetry must be broken below $10^{11}$~GeV.
Accelerated unification can naturally explain the observed gauge
coupling unification by physics below the conformal breaking scale.
In this framework, there is no gravitino and no reason for the existence
of gravitational moduli, thus eliminating the
cosmological problems associated with these particles.
No special dynamics is required to break supersymmetry;
rather, supersymmetry is broken at observable energies because the fixed
point is never reached.
In 4D language, this can be due to irrelevant supersymmetry breaking
operators with approximately equal dimensions.
In 5D language, the size of the extra dimension is stabilized
by massive bulk fields.
No small input parameters are required to generate a large
hierarchy.
Supersymmetry can be broken in the visible sector either through direct
mediation or by the $F$ term of the modulus associated with the
breaking of conformal invariance.
\end{abstract}

\end{titlepage}

\section{Introduction}
If supersymmetry (SUSY) solves the hierarchy problem,
it implies the presence of a new spacetime symmetry in nature.
How does this symmetry arise?
The standard paradigm is that
SUSY is an exact symmetry in the fundamental UV theory,
and is broken spontaneously in the IR.
In this paper, we consider the alternative possibility that 
UV physics is completely non-supersymmetric, and SUSY emerges
as an accidental symmetry in the IR.%
\footnote{The idea that `fundamental' symmetries such as Lorentz invariance
might arise as accidental symmetries of 
IR fixed points has been previously
considered by H.B. Nielsen and collaborators \cite{HBN}.}

An accidental symmetry arises when the UV theory has no relevant
or marginal operators that can be added to the lagrangian to
break the symmetry.
In that case, all symmetry breaking effects flow to zero in
the IR, and the theory becomes invariant under the symmetry at low energies
even if the fundamental theory violates the symmetry maximally.
A famous example is baryon and lepton number in the
standard model.
However, in weakly-coupled theories with scalars, scalar mass
terms are always relevant, so SUSY cannot emerge as an accidental symmetry in such theories.

The situation can be very different in strongly-coupled
theories.
Suppose that there exists a strongly coupled
superconformal theory without any relevant or marginal
SUSY breaking operators that can be added to the lagrangian.%
\footnote{Every 4D CFT has a conserved stress energy operator $T_{\mu\nu}$.
The traceless part of $T_{\mu\nu}$ has dimension 4, but the trace $T$
has an anomalous dimension. 
$T$ is a SUSY breaking operator that can be added to the lagrangian, 
but we assume that it has a positive anomalous dimension so that it
is irrelevant.}
This fixed point will be attractive to all perturbations,
so the boundary of the basin of attraction of the fixed point
will consist of theories that have no approximate SUSY.
For example, $\scr{N} = 1$ $SU(N)$
SUSY QCD with $F$ flavors has a strongly-coupled fixed point
near the middle of the conformal window ($F \simeq 2N$) \cite{Seiberg}.
The scalar masss operators have an uncalculable scaling dimension that is known
to be larger than the canonical dimension \cite{LR}.
It is possible that the anomalous dimensions are large enough
that scalar masses are irrelevant in this theory,
in which case this theory with the addition of large scalar
mass terms flows to a superconformal fixed point in the IR.

Of course, SUSY must be broken at low energies to account for
the absence of superpartners of the observed particles.
The standard paradigm of exact SUSY in the UV requires special structure
to break SUSY in the observable
sector near the weak scale, \eg\ dynamical SUSY breaking.
While many models of dynamical SUSY breaking are known (see \Ref{DSB}
for a review), these are very special theories and SUSY breaking is
generally not robust against perturbations.
The present framework offers an alternative in which
the small scale of SUSY breaking in the visible
sector is simply explained by the fact that
the superconformal fixed point is not reached at low energies.
SUSY is therefore broken explicitly rather than spontaneously;
there is no Goldstino.
There are several possible mechanisms that can prevent the
approach to the fixed point.
One possibility is that there is a relevant SUSY breaking operator
with a small coefficient.
This does not give an explanation of the smallness of the
SUSY breaking, although the small parameter may be natural
if the relevant term breaks a symmetry.
In this paper, we consider the alternative possibility that
the approach to the fixed point is prevented by
\emph{irrelevant} operators.
This is very natural in superconformal theories that have a moduli
space of vacua in the SUSY limit.
In such theories, irrelevant SUSY breaking effects will generate
a potential on the moduli space, and can stabilize the moduli
away from the origin.
If the two lowest-dimension operators have dimensions that
are somewhat close together, this can stabilize the scale modulus
(dilaton) at a scale that is exponentially small compared
to the fundamental scale.
This mechanism is very generic,
and can naturally generate a large hierarchy
without small input parameters or fine tuning.

In this framework, the low-energy
degrees of freedom of the visible sector
are the remnants of the superconformal sector below the 
conformal breaking scale $\La_{\rm IR}$,
which is therefore the compositeness scale
for the standard model matter and gauge particles.
Since SUSY breaking is communicated to the visible sector only
through irrelevant operators, SUSY is an approximate
(but not exact) symmetry of the visible sector, \ie 
SUSY breaking in the visible sector
is naturally below the scale $\La_{\rm IR}$.
Direct mediation of SUSY breaking gives rise to
scalar masses, gaugino masses, $A$ terms, and $\mu$ and $B\mu$ terms,
all of the same size.
Understanding the absence of squark mixing requires additional
structure, as in minimal supergravity.
An alternative is that the CFT dynamics generates an $F$ term
for the dilaton, the modulus associated with the scale of conformal
symmetry breaking.
This naturally gives SUSY breaking below the scale $\La_{\rm IR}$
provided that the CFT has a small parameter that breaks $U(1)_R$
symmetry.
In this case, SUSY breaking in the visible sector is naturally dominated
by anomaly-mediated SUSY breaking \cite{AMSB}.
If the visible sector is the minimal supersymmetric standard model,
the slepton mass-squared terms are negative.
However, completely realistic non-minimal models can be constructed
in this framework, using \eg the ideas of \Refs{PR,NW}.

An important general consequence of this framework is that the
gravitational sector is completely non-supersymmetric.
In particular, there is no gravitino in the spectrum.
This is similar to recent models in which the gravitino
mass is far above the weak scale
\cite{SUSYwoSUGRA,almostnoscale}
(see also \Ref{PG}).
In fact, the present framework can be thought of as a limit of
the model of \Ref{SUSYwoSUGRA} with a high SUSY breaking scale.
The absence of the gravitino eliminates the constraint on the
inflationary reheat temperature that comes from the condition that
gravitinos are not overproduced.
Furthermore, since fundamental physics is non-supersymmetric,
there there are no gravitational moduli, scalar fields with
Planck-suppressed couplings that are
generally present in string theory and higher-dimensional
SUSY theories.
This is a very good thing, because gravitational moduli
cause severe cosmological difficulties that cannot be
`inflated away', and have been viewed as a major obstacle to
realistic string model-building (see \eg \cite{BKN}).
The present models do have a dilaton modulus in the 4D CFT description,
but this modulus
couples with more than gravitational strength, and does not give rise to
cosmological difficulties.

Because gravity is not supersymmetric, gravity loops will generate
SUSY breaking in the visible sector.
However, these loops will be cut off at the scale $\La_{\rm IR}$
where the conformal symmetry and SUSY are restored.
Above the scale $\La_{\rm IR}$, the gravity loops generate a
perturbation corresponding to an irrelevant
operator, which is therefore suppressed by the superconformal dynamics
\cite{LS3}.
In order for the visible scalar masses to be naturally
of order $100\GeV$, the scale
of conformal symmetry breaking must be below $10^{11}\GeV$.

Because the standard model matter and gauge fields are
composite below $10^{16}\GeV$, gauge coupling
unification cannot take place in the usual way.
However, accelerated unification \cite{accelunif} can easily
lower the unification scale below the compositeness scale,
thus explaining the observed unification of the standard model
gauge couplings.

All of the important features of this model arise as direct
consequences of strong conformal dynamics with no relevant operators.
It is now well understood that 5D gravity theories in
anti de Sitter (AdS) space provide `dual' descriptions of 
4D strongly-coupled conformal field theories \cite{ADSCFT, ADSCFTRS}.%
\footnote{5D AdS theories are invariant under $SO(4,2)$, the
4D conformal symmetry,
so it is rigorously true that any 5D AdS theory describes
some 4D conformal field theory.}
We can therefore write explicit 5D models that realize the
framework described above.
The 5D models are of the Randall--Sundrum (RS) type \cite{RS2},
where the UV brane breaks SUSY, while the bulk
and the IR brane are supersymmetric.
The visible sector is localized on the IR brane.
The 5D description is weakly coupled, making explicit calculations
possible.
In particular, we can easily understand SUSY breaking in
the visible sector.
We will construct an explicit 5D model as an existence proof,
but we stress that the main features are generic to superconformal
theories with only irrelevant SUSY breaking operators.

This paper is organized as follows.
In section 2, we present an explicit 5D RS model that realizes the
ideas outlined above.
We consider radius stabilization and construct the low-energy
4D effective field theory, which we use to analyze SUSY breaking
in the visible sector.
We interpret our results in the language of 4D conformal field
theories and argue that the basic features are very general.
In section 3, we briefly discuss phenomenology, and section 4
contains our conclusions.

\section{5D Model}
\subsection{Definition of the Model}
Our model is based on the Randall--Sundrum (RS) model \cite{RS2}.
This is a 5D spacetime compactified on a $S^1 / Z_2$ orbifold,
with metric
\beq[RSmetric]
ds^2 = e^{-2 \si(y)} \eta_{\mu\nu} dx^\mu dx^\nu
+ d y^2,
\eeq
where $y$ is a periodic variable with period $2\ell$,
and $\si(y)$ is a periodic function defined by
\beq
\si(y) = k |y|
\ \ \hbox{\rm for}\  -\ell < y \le +\ell.
\eeq
Note that this implies
\beq
\si' &= \frac{d\si}{d y} = k \sgn(y)
\ \ \hbox{\rm for}\  -\ell < y \le +\ell,
\\
\si'' &= \frac{d^2 \si}{d y^2}
= 2 k \left[ \de(y) - \de(y - \ell) + \cdots \right].
\eeq
The physical region is $0 \le y \le \ell$.
The boundary at $y = 0$ is the `UV brane'
(or `Planck brane') where the zero mode of graviton is localized,
and the boundary at $y = \ell$ is the `IR brane.'
We assume that the physics of this model is controlled by a single
fundamental scale $\LUV \sim M_{\rm P}$.
This means that we take all couplings in the action to be of order
1 in units of $\LUV$.
The effect of the metric \Eq{RSmetric} is that physical
scales on the IR brane are `warped down' to the scale
\beq
\LIR = \LUV \om,
\eeq
where
\beq
\om = e^{-k \ell}
\eeq
is the `warp factor.'
There can be an exponentially large hierarchy between
the fundamental scale and $\LIR$, provided that the size
of the extra dimension $\ell$ can be stabilized at a value somewhat
larger than $k^{-1} \sim \LUV^{-1}$, so that $\om \ll 1$.
%
%
This is the hierarchy generating mechanism of Randall and Sundrum
\cite{RS2}.

The RS model is interesting in its own right, but an additional
motivation to consider this model is that it can be interpreted as
a strongly-coupled 4D conformal field theory (CFT)
\cite{ADSCFTRS,APR,RZ}.
The origin of this equivalence is that the bulk AdS$_5$
geometry has a $SO(4,2)$ symmetry, which is isomorphic to the
4D conformal group, which acts on the branes as a
4D conformal transformation.
In this equivalence, the bulk Kaluza--Klein (KK) modes
are identified with excitations of the CFT.
The UV brane acts as a UV cutoff on these modes, while the
IR brane gives rise to spontaneous breaking of the conformal
invariance.%
\footnote{More precisely, the conformal symmetry is nonlinearly
realized by the position of the
IR brane in the limit where the UV brane is
at infinity \cite{RZ}.}
Bulk fields are associated with operators of the CFT.
Scalar operators that are irrelevant (respectively relevant)
are associated with scalar modes with bulk mass $m^2 > 0$
(respectively $m^2 < 0$).

We are interested in 4D CFT's where the UV physics breaks SUSY,
but the theory flows toward a superconformal fixed point in the IR.
This means we want a 5D model where the couplings on the
UV brane break SUSY maximally,
while the action for the bulk and the
IR brane is supersymmetric.
This is radiatively stable by 5D locality.
We also want the CFT to have only irrelevant perturbations.
This means that all scalar fields must have positive bulk masses.
We will therefore add massive hypermultiplets in the bulk, which
will play an important role in stabilization
and SUSY breaking.%
\footnote{The trace of the CFT stress-energy tensor corresponds to the
5D dilaton state.
We therefore assume that the 5D dilaton is more massive than the
hypermultiplets.}
%
%

At energies below the mass of the lightest KK mode $m_{\rm KK} \sim \LIR$
the physics can be described by a 4D effective theory
that is approximately supersymmetric.
The light degrees of freedom consist of the $\scr{N} = 1$
SUGRA multiplet, the radion chiral multiplet
\beq
\om = e^{-k \ell} + \cdots + \th^2 F_\om,
\eeq
and the light fields localized on the IR brane.
The effective lagrangian is \cite{LS2}
\beq[L4eff]
\bal
\scr{L}_{4,{\rm eff}} &= -\frac{M_5^3}{k} \myint d^4\th\,
(\om^\dagger \om - \varphi^\dagger \varphi)
\\
&\qquad
+ \myint d^4\th\, \om^\dagger \om K_{\rm IR}
+ \left( \myint d^2\th\, \om^3 W_{\rm IR} + \hc \right)
\\
&\qquad
+ \hbox{\rm SUSY\ breaking\ terms},
\eal\eeq
where $\varphi = 1 + \th^2 F_\varphi$ is the conformal compensator
and the superspace integrals are shorthand for the the covariant
$F$ and $D$ projections of the superconformal tensor calculus
\cite{confSUGRA}.
The 4D Planck scale is
\beq[M4]
\MP^2 = \frac{M_5^3}{k} ( 1 - \om^2 )
\simeq \frac{M_5^3}{k}.
\eeq
$K_{\rm IR}$ and $W_{\rm IR}$ are the \Kahler potential and
superpotential of the fields localized on the IR brane.
Note that the field $\om$ has a canonical kinetic term,
and that it couples to physics on the IR brane as a dilaton.
This is the nonlinear realization of the conformal symmetry,
which will play an important role in what follows.

\subsection{SUSY Breaking from 5D SUGRA}
We now begin our discussion of SUSY breaking on the IR brane
(the visible sector).
Any SUSY breaking effects must arise by communication with the UV brane
via bulk modes.
In this subsection, we discuss the effects of the 5D SUGRA fields.
We will discuss the effects of the bulk hypermultiplets after we
have discussed stabilization.

Tree-level SUGRA KK exchange does not give rise to SUSY breaking
operators on the IR brane \cite{LS2,LS1,LLP}.
There is a potential tree-level SUSY breaking effect from a
constant superpotential on the IR brane.
This generates a nonzero VEV for $F_\om$,
which gives rise to
anomaly-mediated SUSY breaking on the IR brane \cite{SUSYwoSUGRA}.
If the constant superpotential is order $1$ in units of $\LUV$,
we obtain
$F_\om / \om \sim \LIR$,
where the \lhs is the order parameter for anomaly mediation on
the IR brane. 
In order to obtain SUSY breaking masses at the weak scale, we must
have $\LIR \lsim 10\TeV$.
Since $\LIR$ is also the compositeness scale for the
standard model gauge fields, this implies that the
standard model gauge fields are strongly
coupled below $10$~TeV.
This requires a large
number of charged states below $10$~TeV, and the masses of these extra
states must be finely tuned to get the observed low-energy gauge
couplings.
To avoid this unattractive scenario, we 
assume that the constant superpotential term is absent
or small,
which is natural by $U(1)_R$ symmetry.

We now consider SUGRA loop contributions to SUSY breaking on
the IR brane, \eg scalar masses.
For loop momenta below 
$m_{\rm KK} \sim \LIR$, the loop diagram is
identical to a 4D loop diagram with a graviton line.
This integral is effectively cut off for momenta above
$m_{\rm KK} \sim \LIR$ by higher-dimensional locality.
This gives  \cite{SUSYwoSUGRA}
\beq[SUGRAloopmass]
\De m^2_{\rm scalar} \sim \frac{1}{16\pi^2}\,
\frac{\LIR^4}{\MP^2}.
\eeq
Demanding that this contribution to the
scalar masses be of order $100\GeV$ or less gives
\beq
\LIR \lsim 10^{11}\GeV.
\eeq
If $\LIR \sim 10^{11}$~GeV, this gives a flavor
universal contribution to the scalar masses of order
100~GeV.
In models where SUSY breaking in the visible sector is anomaly-mediated,
an additional positive scalar mass-squared contribution
can make the slepton masses positive in the minimal supersymmetric
standard model. 
It would therefore be extremely interesting to compute the
sign of the scalar mass contribution \Eq{SUGRAloopmass}.%
\footnote{For the calculation in flat 5D theories, see
\Refs{LLP2,RSS}.}
Gaugino masses and $A$ terms from SUGRA loops are negligibly small.

\subsection{Casimir Energy}
Because SUSY is broken, there will be a nonzero Casimir energy from
states in the bulk that feel SUSY breaking via couplings to the
UV brane.
The Casimir energy depends on the radius, and therefore contributes to the
radius potential.

The Casimir energy can be written as a sum over KK modes.
(In the 4D CFT interpretation, these are bound states of the strongly-coupled
CFT.)
The radius-dependent part of the Casimir energy is UV finite,
and is therefore dominated by the contribution from
the lowest lying KK states, which are approximately supersymmetric.
The SUSY violating mass splittings are due to gravitational strength
interactions, and are therefore of order
\beq
\De m_{\rm KK} \sim \frac{m_{\rm KK}^3}{\MP^2} \sim \om^3.
\eeq
The Casimir energy vanishes when the spectrum is supersymmetric,
so we have
\beq
V_{\rm Casimir} \sim m_{\rm KK}^3 \Delta m_{\rm KK} \sim \om^6.
\eeq
This agrees with explicit calculations (see \eg \Ref{GP}).%
\footnote{We thank A. Pomarol for helpful discussions on Casimir
energy.}
We will consider stabilization mechanisms such that this contribution to
the potential is negligible, so we do not need to know the
sign of the Casimir energy.

\subsection{Radius Stabilization}
We now give a detailed discussion of radius stabilization
in the 5D model outlined above.
The 4D CFT interpretation of the radion is the scale modulus
(dilaton) associated with
spontaneous conformal symmetry breaking \cite{RZ}, so this is
equivalent to dilaton stabilization in the CFT.

In the 5D description, the radion potential is generated by the
Goldberger--Wise mechanism \cite{GW} with
bulk scalar fields with positive mass-squared.%
\footnote{\Ref{GW} considered scalars with $m^2 \ll k^2$,
corresponding to almost marginal 4D CFT operators;
we consider $m^2 \gsim k^2$.}
In the CFT description, the bulk scalars parameterize the effects
of irrelevant operators in the CFT.
For a single scalar, we expect a potential of the form
\beq
V_{\rm eff} = a \om^n,
\eeq
with $n > 0$.
By itself this will give a runaway potential, but we can
obtain a stable minimum if there are several such terms:
\beq[Vracetrack]
V_{\rm eff} = a_1 \om^{n_1} + a_2 \om^{n_2},
\eeq
with $n_1 > n_2 > 0$.
We find a local minimum at a small value of $\om$
if $n_1$ and $n_2$ are close in value.
If we define $\ep = n_1 - n_2$, then
\beq
\om = \left( - \frac{(n_1 - \ep) a_2}{n_1 a_1} \right)^{1/\ep}
\eeq
is a minimum provided that $a_1 > 0$, $a_2 < 0$.
(In fact, $V_{\rm eff} < 0$ at the minimum, so this vacuum has lower than
the asymptotic vacuum $\om = 0$.)
Note that $\om$ is naturally exponentially small if
the factor in parentheses is positive and less than one,
and $\ep$ is moderately small.
The potential \Eq{Vracetrack} has the same form as the modulus potential
in `racetrack' models \cite{racetrack}.
The radion field $\om$ has a kinetic term with coefficient
$\MP^2$ (see \Eqs{L4eff} and \eq{M4}), so the physical
mass of the radion is of order
\beq[mradion]
m^2_{\rm radion} \sim \frac{\ep a_1}{\MP^2} \om^{n_1 - 2}.
\eeq

We now discuss in detail how potentials of this form can arise
from bulk hypermultiplets.
The action for a hypermultiplet in the RS background
was given in terms of $\scr{N} = 1$ superfields
by Mart\'\i{} and Pomarol in \Ref{MP}.
We add a general SUSY breaking potential on the UV brane,
and a superpotential on the IR brane.
These are the leading brane-localized terms in a low-energy expansion.
The action is therefore
\beq
S_{\rm hyper} = \myint d^4 x \int_{-\ell}^\ell d y\, \scr{L}_5,
\eeq
where%
\footnote{We write the action in terms of the two-sided derivative
$\Pt \dd_y \Phi = \Pt \d_y \Phi - (\d_y \Pt) \Phi$.
This is without loss of generality, since
$f \Pt \dd_y \Phi = 2 f \Pt \d_y \Phi + \d_y f \Pt \Phi
+ \hbox{\rm total\ derivative}$}
\beq[HyperL]
\bal
\scr{L}_5 &= \myint d^4\th\,e^{-2\si} (\Phi^\dagger \Phi + \Pt^\dagger \Pt)
+ \left[ \myint d^2\th\, e^{-3\si} \left(
\sfrac 12 \Pt \dd_y \Phi + c \si' \Pt \Phi \right) + \hc \right]
\\
& \qquad
- \de(y) U(\Phi, F)
+ \de(y - \ell)\,  \om^3 \left[ \myint d^4\th\, 
W(\Phi) + \hc \right].
\eal\eeq

The AdS/CFT correspondence relates the mass of states in the bulk to the
dimension of an operator in the CFT.
A 5D scalar with bulk mass $m$ corresponds to an operator of dimension
$d = 2 + \sqrt{4 + m^2 / k^2}$.
The bulk mass of the scalars from the hypermultiplet action above is
\beq
m_{\Phi,\Pt}^2 = k^2 (c \mp \sfrac 32)(c \pm \sfrac 52).
\eeq
so the dimensions of the operators associated with the scalar
components of $\Phi$ and $\Pt$ are
\beq
\dim(\scr{O}_{\Phi, \Pt}) = 2 + | c \pm \sfrac 12 |.
\eeq
If we want the operators associated with both $\Phi$ and $\Pt$ to be
irrelevant, we must have $|c| > \sfrac 52$.

For a scalar $\phi$ of mass $m$ corresponding to an operator of dimension
$d$, the general solution to the bulk equations of motion is
\beq[scalarsoln]
\phi = A e^{d\si} + B e^{(4 - d)\si}.
\eeq
The coefficients $A$ and $B$ are fixed by the boundary conditions.
The CFT interpretation of the coefficients is as follows.
$A$ is associated with a VEV of the operator
\beq
A = \frac{\avg{\scr{O}}}{2 d - 4},
\eeq
while $B$ is associated with adding a term to the CFT lagrangian
\beq
\De\scr{L}_{\rm CFT} = \la \scr{O},
\eeq
with
\beq
B \propto \la.
\eeq
For irrelevant operators ($d > 4$) the second term in
\Eq{scalarsoln} is exponentially
decreasing in the IR, and therefore the value of the coefficient $B$
will be determined by physics on the UV brane.
In models where all dimensionful couplings are order 1 in units of
$\LUV$, we therefore expect $B \sim \LUV^{3/2}$.
The first term in \Eq{scalarsoln}
grows in the IR, and is therefore determined by physics on the
IR brane.
The scale of physics on the IR brane is set by $\LIR$,
so this will generally fix $A \sim \LIR^{3/2} \ll \LUV^{3/2}$.
We therefore say that the first (second) term in \Eq{scalarsoln} is IR (UV)
dominated, respectively.

We now solve the equations of motion.
We look for solutions depending only on $y$.
The $\Pt$ equation of motion is
\beq[Hteom]
\d_y F + (c - \sfrac 32) \si' F = 0.
\eeq
Imposing the condition that $F$ is periodic and even, the most general
solution is
\beq[Fsoln]
F = F_0 e^{-(c - \frac 32) \si},
\eeq
where $F_0$ is a constant of integration.
We can define
\beq
F_{\rm UV} &= F(0) = F_0,
\\
F_{\rm IR} &= F(\ell) = F_0 \om^{c - \frac 32}.
\eeq

The $\Phi$ equation of motion is
\beq[Heom]
e^{-3\si} \d_y \Ft - (c + \sfrac 32) \si' e^{-3\si} \Ft
= -\de(y) \frac{\d U}{\d \Phi} + \de(y - \ell) \om^3 
\frac{\d^2 W}{\d \Phi^2} F.
\eeq
Imposing the condition that $F$ is periodic and odd, the most general
solution is
\beq[Ftsoln]
\Ft = \Ft_0 \frac{\si'}{k}  e^{(c + \frac 32) \si},
\eeq
where $\Ft_0$ is a constant of integration.
Even though $\Ft$ is odd under the orbifold $Z_2$, it is discontinuous
at the boundaries, so it effectively has a nonvanishing value on each
boundary.
It is convenient to define
\beq
\tilde{F}_{\rm UV} &= \lim_{y \to 0+} \Ft = \Ft_0,
\\
\tilde{F}_{\rm IR} &= \lim_{y \to \ell-} \tilde{F}
= \Ft_0 \om^{-(c + \frac 32)}.
\eeq
In terms of these quantities, the jump conditions
at the UV and IR branes are
\beq[Ftjump]
\Ft_{\rm UV} &= -\frac 12 \frac{\d U}{\d \Phi_{\rm UV}},
\\
\eql{Ftjump2}
\Ft_{\rm IR} &= -\frac 12 \frac{\d^2 W}{\d \Phi_{\rm IR}^2} F_{\rm IR},
\eeq
where
$\Phi_{\rm UV} = \Phi(0)$,
$\Phi_{\rm IR} = \Phi(\ell)$.

The $\Ft$ equation of motion is
\beq[Fteom]
e^{-3\si} \d_y \Phi + (c - \sfrac 32) \si' e^{-3 \si} \Phi
+ e^{-2\si} \Ft^\dagger = 0.
\eeq
Imposing the condition that $\Phi$ is periodic and even,
the most general solution is
\beq[Hsoln]
\Phi = \Phi_0 e^{-(c - \frac 32) \si}
- \frac{\Ft_0^\dagger}{(2c + 1) k} e^{(c + \frac 52) \si},
\eeq
where $\Phi_0$ is a constant of integration.

Finally, the $F$ equation of motion is
\beq[Feom]
e^{-3\si} \d_y \Pt - (c + \sfrac 32) \si' e^{-3\si} \Pt 
- e^{-2\si} F^\dagger = -\de(y) \frac{\d U}{\d F}
+ \de(y - \ell) \om^3 \frac{\d W}{\d \Phi}.
\eeq
Imposing the condition that $\Pt$ is periodic and odd,
the most general solution is
\beq[Htsoln]
\Pt = \frac{\si'}{k} \left[
\Pt_0 e^{(c + \frac 32) \si}
- \frac{F_0^\dagger}{(2 c - 1) k} 
e^{-(c - \frac 52) \si} \right],
\eeq
where $\Pt_0$ is a constant of integration.
To write the boundary conditions, we define
\beq
\Pt_{\rm UV} &= \lim_{y \to 0+} \Pt = \Pt_0 - \frac{F_0^\dagger}{(2c - 1) k},
\\
\Pt_{\rm IR} &= \lim_{y \to \ell-} \Pt = \Pt_0 \om^{-(c + \frac 32)}
- \frac{F_0^\dagger}{(2 c - 1) k} \om^{c - \frac 52}.
\eeq
The jump conditions at the UV and IR branes can then be written
\beq[Htjump]
\Pt_{\rm UV} &= -\frac 12 \frac{\d U}{\d F_{\rm UV}},
\\
\eql{Htjump2}
\Pt_{\rm IR} &= -\frac 12 \frac{\d W}{\d \Phi_{\rm IR}}.
\eeq

The fields
$\Pt$ and $\Ft$ are discontinuous at the boundaries.
In our formulation with auxiliary fields, this is because
the equations are first-order with delta function terms on the
boundaries.
Integrating out the auxiliary fields give rise to terms proportional
to powers of delta functions, which na\"\i{}vely are too singular to have
a good continuum limit.
However, supersymmetry and the orbifold projection
evidently make sense out of these singular brane terms
and give rise to the discontinuities at the boundaries.

To summarize, the solution is given by \Eqs{Fsoln}, \eq{Ftsoln},
\eq{Hsoln}, \eq{Htsoln}.
The four constants of integration $F_0$, $\Ft_0$, $\Phi_0$, and
$\Pt_0$ are to be determined from the four discontinuity conditions in
\Eqs{Ftjump}, \eq{Ftjump2}, \eq{Htjump}, and \eq{Htjump2}.
Note that the jump conditions 
contain no explicit dependence on $\om$ when expressed entirely
in terms of UV or IR quantities.

We now use these results to write the effective potential.
At the classical level, this is obtained by substituting the
solutions to the equations of motion given above into the action
and performing the
integral over $y$ to obtain a potential that depends on $\om$.
Because the bulk terms are quadratic in $\Phi$ and $\Pt$, imposing the
bulk equations reduces the effective potential to boundary terms.
Using the jump conditions to simplify the result,
we obtain the rather elegant expression
\beq[Veff]
\bal
V_{\rm eff} = U(\Phi_{\rm UV}, & F_{\rm UV})
+ \left( \Phi_{\rm UV} \Ft_{\rm UV} + \Pt_{\rm UV} F_{\rm UV}
+ \hc \right)
\\
&+ \om^3 \left( -\Phi_{\rm IR} \Ft_{\rm IR}
+ \Pt_{\rm IR} F_{\rm IR} + \hc \right).
\eal\eeq
There is implicit dependence on $\om$ through the
boundary values of the fields. 

We now specialize to the case where $c > \sfrac 52$, so that
the scalar components of $\Phi$ and $\Pt$ are associated with
operators of dimension
\beq
d = \dim(\scr{O}_\Phi) = c + \sfrac 52,
\qquad
\tilde{d} = \dim(\scr{O}_{\Pt}) = c + \sfrac 32,
\eeq
with $d, \tilde{d} > 4$.
We also assume that all couplings in the lagrangian
are order one in units of $\LUV$.
Then we have
\beq
F = F_{\rm UV} e^{(4 - d) \si}.
\eeq
This is UV dominated, so we parameterize it by $F_{\rm UV}$.
We also have
\beq
\Ft = \frac{\si'}{k} \,
\Ft_{\rm IR} \om^{\tilde{d}} e^{\tilde{d} \si}.
\eeq
This is IR dominated, so we parameterize it by $\Ft_{\rm IR}$.
The solution for $\Phi$ is
\beq[Phisoln]
\Phi = \left[ \Phi_{\rm UV}
+ \frac{\Ft_{\rm IR}^\dagger \om^{\tilde{d}}}{(2d - 4) k} \right]
e^{(4 - d)\si}
- \frac{\Ft_{\rm IR}^\dagger \om^{\tilde{d}}}{(2d - 4) k} e^{d \si}.
\eeq
We will see that $\Ft_{\rm IR} \ll \scr{O}(\om)$ as a result of the jump equations,
so the last term is small for all values of $y$.
$\Phi$ is therefore UV dominated, and we parameterize it
using $\Phi_{\rm UV}$.
Finally, we have
\beq
\Pt = \frac{\si'}{k} \left[
\left( \Pt_{\rm IR} \om^{\tilde{d}}
+ \frac{F_{\rm UV}^\dagger}{(2 \tilde{d} - 4) k} \om^{2 \tilde{d} - 4} \right)
e^{\tilde{d} \si}
- \frac{F_{\rm UV}^\dagger}{(2 \tilde{d} - 4) k} e^{(4 - \tilde{d})\si} \right].
\eeq
$\Pt$ is unsuppressed at both the UV and IR branes.
We parameterize it by $\Pt_{\rm IR}$.

The constants of integration
$F_{\rm UV}$, $\Ft_{\rm IR}$, $\Phi_{\rm UV}$, and $\Phi_{\rm IR}$
are determined by the jump equations.
\Eqs{Ftjump} and \eq{Ftjump2} are
\beq[Ftjumpnew]
\Ft_{\rm IR} &= - \frac 12 F_{\rm UV} \om^{d - 4}
\frac{\d^2 W}{\d \Phi_{\rm IR}^2} , 
\\
\eql{Ftjumpnew2}
\frac{\d U}{\d \Phi_{\rm UV}} &= -2 \Ft_{\rm IR} \om^{\tilde{d}}.
\eeq
\Eq{Ftjumpnew} implies that
$\Ft_{\rm IR} \lsim \scr{O}(\om^{d - 4}) \ll \scr{O}(\om)$,
implying that $\Phi$ is UV dominated (see \Eq{Phisoln}).
The jump conditions \Eqs{Htjump} and \eq{Htjump2} can be written
\beq[Htjumpnew]
F_{\rm UV}^\dagger
&= \frac{(2 \tilde{d} - 4) k}{1 - \om^{2 \tilde{d} - 4}} \left[
\frac 12 \frac{\d U}{\d F_{\rm UV}}
+ \Pt_{\rm IR} \om^{\tilde{d}} \right],
\\
\eql{Htjumpnew2}
\Pt_{\rm IR} &= - \frac 12 \frac{\d W}{\d \Phi_{\rm IR}}.
\eeq
From \Eqs{Ftjumpnew} through \eq{Htjumpnew2} we can see that
there are generically solutions with
$\Phi_{\rm UV}, F_{\rm UV} = \scr{O}(\om^0)$.
The leading approximation for $\Phi_{\rm UV}$ and $F_{\rm UV}$
is obtained by solving
\beq
0 &= \frac{\d U}{\d \Phi_{\rm UV}},
\\
F_{\rm UV}^\dagger &= (\tilde{d} - 2) k \frac{\d U}{\d F_{\rm UV}}.
\eeq
The corrections depend on the form of the IR superpotential.

Let us consider an example where
\beq
W = \ka \Phi.
\eeq
We then have
\beq
\Pt_{\rm IR} = -\sfrac 12 \ka,
\qquad
\Ft \equiv 0,
\eeq
while $\Phi_{\rm UV}$ and $F_{\rm UV}$ are determined by solving
the equations
\beq[finaljump1]
0 &= \frac{\d U}{\d \Phi_{\rm UV}},
\\
\eql{finaljump2}
F_{\rm UV}^\dagger &= \frac{(\tilde{d} - 2) k}
{1 - \om^{2 \tilde{d} - 4}}\,
\left( \frac{\d U}{\d F_{\rm UV}} - \ka \om^{\tilde{d}} \right).
\eeq
Expanding in powers of $\om$,
\beq\bal
\Phi_{\rm UV} &= \Phi_{\rm UV}^{(0)} + \om^{\tilde{d}} \Phi_{\rm UV}^{(1)}
+ \scr{O}(\om^{2 \tilde{d} - 4}),
\\
F_{\rm UV} &= F_{\rm UV}^{(0)} + \om^{\tilde{d}} F_{\rm UV}^{(1)}
+ \scr{O}(\om^{2 \tilde{d} - 4}),
\eal\eeq
we obtain the leading $\om$-dependent contribution to the
effective potential:
\beq
V_{\rm eff} = \left[
- \ka F_{\rm UV}^{(0)} + \hc \right] \om^{\tilde{d}}
+ \scr{O}(\om^{2\tilde{d} - 4}).
\eeq
It is clear that the coefficient of $\om^{\tilde{d}}$
can have either sign, as required for
stabilization.
If we want the Casimir energy to be negligible compared to the effects
discussed here, we must have
$\tilde{d} < 6$.

The results above are in agreement with the expectations of the
AdS/CFT correspondence.
The CFT operator of lowest dimension associated with the hypermultiplet
is $\scr{O}_{\Pt}$, with dimension
$\tilde{d}$.
If we add to the CFT lagrangian a term
\beq
\De \scr{L}_{\rm UV} = \tilde\la \scr{O}_{\Pt},
\eeq
conformal invariance implies that 
the effective potential has the form
\beq
V_{\rm eff} = \LIR^4 f(\tilde{\la}_{\rm eff}(\LIR)),
\eeq
where $\tilde{\la}_{\rm eff}(\mu) \sim \mu^{\tilde{d} - 4}$
is the renormalized effective coupling at the scale $\mu$.
The vacuum energy vanishes for $\la = 0$ by SUSY, so $f(0) = 0$.
Expanding in powers of $\tilde\la$ therefore gives (using $\LIR \sim \om$)
\beq
V_{\rm eff} \sim \om^{\tilde{d}} + \om^{2 \tilde{d} - 4} + \cdots,
\eeq
which agrees
with the 5D result found above.

We see that we can obtain a potential of the form
\Eq{Vracetrack} with the addition of two hypermultiplets with mass
parameter $c > \sfrac 52$.%
\footnote{
Alternatively, we could use one one hypermultiplet with
$c \simeq \sfrac 92$ (corresponding to $\tilde{d} \simeq 6$)
together with Casimir energy.}
To obtain a large hierarchy, we need the masses of the two hypermultiplets
to be approximately equal, but the tuning required is only logarithmic.

\subsection{SUSY Breaking from Stabilization}
We now consider the size of SUSY breaking on the IR
brane arising from the radius stabilization dynamics.

The hypermultiplet $F$ terms can give rise to direct
SUSY breaking from couplings on the IR brane of the form
(in units where $\LUV = 1$)
\beq[directSUSY]
\bal
\De \scr{L}_{\rm IR} \sim & \myint d^4 \th\, \Phi^\dagger \Phi
\left( Q^\dagger Q + H^\dagger H \right)
+ \left( \myint d^2\th\, \Phi W^\al W_\al + \hc \right)
\\
& \qquad
+ \myint d^2 \th\, \Phi Q^2 H + \hc
\\
& \qquad
+ \myint d^4\th \left( \Phi^\dagger H^2 + \Phi^\dagger \Phi H^2
+ \hc \right),
\eal\eeq
where $Q$ and $H$ are matter and Higgs fields localized on the IR brane,
and $W_\al$ is the field strength for standard model gauge fields, also
localized on the IR brane.
Just as in `minimal SUGRA' models, this gives rise to scalar masses, gaugino
masses, $A$ terms, as well as $\mu$ and $B\mu$ terms
naturally of the same
size, namely
\beq[MSUSY]
M_{\rm SUSY} \sim F_{\rm IR} \sim \om^d \sim \La_{\rm IR} \om^{\tilde{d} - 4},
\eeq
where we have restored the mass scales in the last step.
Note that $M_{\rm SUSY} \ll \LIR$, as expected.
As a mechanism for SUSY breaking, this has the attraction of simplicity
and elegance;
in particular, it generates $\mu$ and $B\mu$ terms without
additional complicated structure \cite{GM}.
There is however no explanation of the absence of squark mixing required to
avoid large flavor-changing neutral currents.
This type of SUSY breaking therefore requires additional flavor structure
at high scales, such as the models of \Refs{SUSYflavor}.

In the 4D CFT interpretation, the effects parameterized
by \Eq{directSUSY} represent SUSY breaking effects of the composite
CFT states from irrelevant CFT operators.
Like the operators in the 5D description, these are very generic
effects that are expected to be present unless there are special
symmetries that forbid them.
It is remarkable that these effects can naturally give rise to all required
SUSY breaking at the same scale.

Another potentially important source of SUSY breaking is the radion $F$ term
generated from the stabilization dynamics.
The stabilization breaks SUSY, as can be seen from the nonzero value
of the hypermultiplet $F$ terms in the bulk.
To compute the radion $F$ term, we will use the technique of analytic
continuation into superspace.
The strategy is to write all SUSY breaking terms in the lagrangian
using superfield spurions, and obtain a superfield form of the 4D
effective potential.
We can then find the dependence on the radion $F$ term in the 4D
effective theory by promoting
$\om$ to a chiral superfield \cite{LS2}.

We begin by writing the SUSY-breaking potential on the UV brane
in terms of a superfield spurion:
\beq[HyperL]
\De \scr{L}_5 = \de(y) \myint d^4\th\, S(\Phi, \De \Phi),
\eeq
where
\beq
S(\Phi, F) = -\th^4\, U(\Phi, F),
\eeq
and we use the abbreviations
\beq[Deabbrev]
\De = -\sfrac 14 D^2,
\qquad
\bar\De = -\sfrac 14 \bar{D}^2.
\eeq
The $\Pt$ and $\Phi$ superfield equations of motion are
\beq
e^{-2\si} &\bar{\De} \Pt^\dagger
+ e^{-3 \si} \left[ \d_y \Phi + (c - \sfrac 32) \si' \Phi \right]
= 0,
\\
e^{-2\si} &\bar\De \Phi^\dagger
+ e^{-3 \si} \left[ -\d_y \Pt + (c + \sfrac 32) \si' \Pt \right]
\nonumber\\
& \qquad
= - \de(y) \bar\De \left[
\frac{\d S}{\d \Phi} + \De \left( \frac{\d S}{\d F} \right)
\right]
- \de(y - \ell) \om^3 \frac{\d W}{\d \Phi}.
\eeq
The solutions are
\beq
\Phi &= \Phi_0 e^{-(c - \frac 32) \si}
- \frac 1{(2c + 1)k} \bar\De \Pt_0^\dagger
e^{(c + \frac 52) \si},
\\
\Pt &= \frac{\si'}{k} \left[
\Pt_0 e^{(c + \frac 32) \si} - \frac 1{(2c - 1)k}
\bar\De \Phi_0^\dagger e^{-(c - \frac 52)\si} \right],
\eeq
where $\Phi_0$ and $\Pt_0$ are superfield constants of integration,
determined by the jump equations on the UV and IR branes:
\beq
\Pt_0 - \frac{1}{(2c - 1) k}
\bar\De \Phi_0^\dagger 
&= \sfrac 12\bar\De \left[
\frac{\d S}{\d \Phi} + \De \left( \frac{\d S}{\d F}
\right) \right]_{\rm UV},
\\
\Pt_0 \om^{-(c - \frac 32)}
- \frac{1}{(2c - 1) k} \bar\De \Phi_0^\dagger \om^{c - \frac 52}
&= -\sfrac 12 \left[ \frac{\d W}{\d \Phi} \right]_{\rm IR}.
\eeq
The effective potential is
\beq[VeffSS]
\bal
V_{\rm eff} &= \myint d^4\th \left[
-S + \sfrac 12 \left(
\Phi \frac{\d S}{\d \Phi} + F \frac{\d S}{\d F} + \hc \right) \right]_{\rm UV}
\\
&\qquad
+ \myint d^2\th\, \om^3 \left[ -W + \sfrac 12 \Phi \frac{\d W}{\d \Phi}
\right]_{\rm IR} + \hc
\eal\eeq
Note that this depends on $\om$ implicitly via the solutions for
$\Phi_0$ and $\Pt_0$.
It is easily verified that these expressions reproduce the component
expressions given earlier.

The expression \Eq{VeffSS} can be directly analytically continued into
superspace by promoting $\om$ to a chiral superfield.
To compute the VEV of $F_\om$, we need the linear term in $F_\om$ in
the effective potential.
(The leading quadratic term in $F_\om$ comes from \Eq{L4eff}.)
Note that there is no contribution to the coefficient of $F_\om$ from
the first term in \Eq{VeffSS}, which is a function of UV quantities.
Even though this term depends implicitly on $\om$ through $\Phi_0$, all
of the $\th$ integrations must act on the explicit $\th^4$ in $S$ to
get a nonzero result.
This immediately guarantees that $F_\om / \om \lsim \LIR$, and
agrees with the expectation that physics associated with the UV brane
does not generate an $F$ term for the radion.
A similar argument shows that the Casimir energy from bulk SUGRA does not
generate a coefficient for $F_\om$.

There is a nonzero linear term for $F_\om$
from the second term in \Eq{VeffSS}, which depends
on IR quantities.
This is
\beq[Fomcoeff]
\bal
V_{\rm eff} = \om^2 \Biggl[ &
3 \left( -W + \sfrac 12 \Phi \frac{\d W}{\d \Phi} \right)
\\
&
+ \sfrac 12 \om \frac{\d \Phi_{\rm IR}}{\d \om}
\left( -\frac{\d W}{\d \Phi} + \Phi \frac{\d^2 W}{\d \Phi^2} \right)
\Biggr]_{\rm IR}
F_\om + \hc + \cdots
\eal\eeq
Note that the coefficient of $F_\om$ vanishes identically if $W = \la \Phi^2$.
This makes sense, since this term
preserves conformal symmetry ($\la$ is dimensionless). 

For the example considered above, $W = \ka \Phi$
and $\Phi_{\rm IR} \sim \om^{d - 4}$,
we have
\beq
\frac{F_\om}{\om}
\sim \om \Phi_{\rm IR} \sim \La_{\rm IR} \om^{d - 4}.
\eeq
The \lhs is the order parameter for anomaly mediated SUSY breaking (AMSB) on the
IR brane \cite{SUSYwoSUGRA}.
Comparing to \Eq{MSUSY}, we see that AMSB from
this source is always smaller than direct SUSY breaking, even without taking
into account the loop suppression factors in AMSB.

Integrating out $F_\om$ also generates SUGRA corrections to the radion
potential.
This gives
$\De V_{\rm eff} \sim \om^{2 d - 4}$,
which is negligible compared to the $\om^{\tilde{d}}$ term found above.

We can obtain an additional contribution to the VEV of
$F_\om / \om$ if there is a small constant term in the superpotential
on the IR brane:
\beq
\De W = C
\eeq
gives
\beq
\frac{F_\om}{\om} \sim \La_{\rm IR} \left( \frac{C}{\La_{\rm UV}^3} \right).
\eeq
As discussed above, if there are no small parameters in the theory, we
have $C \sim \La_{\rm UV}^3$ and the IR scale is too low.
However $C \ll \La_{\rm UV}^3$ is natural because $C$ breaks a $U(1)_R$
symmetry.
In the 4D CFT interpretation, the CFT has a small parameter that breaks
the $U(1)_R$ symmetry and dynamically generates a small $F$ term for the
dilaton.
In this scenario, SUSY breaking can be dominated by AMSB.
This is attractive because it automatically gives flavor-blind SUSY breaking.

\section{Phenomenology and Cosmology}
In this section, we make some brief remarks about phenomenology
and cosmology.

First we consider the radion, which is the only model-independent new
degree of freedom in this framework.
The radion is localized near the IR brane, and so its couplings to visible
matter are suppressed by powers of $\LIR$ rather than being Planck
suppressed.
The corresponding mode in 4D CFT language is the dilaton, which is a
bound state of the CFT dynamics at the scale $\LIR$.
This modulus therefore couples more strongly than gravitational moduli,
making the cosmology much safer.
Because the radion decouples in the conformal limit, its couplings will be
suppressed by loop factors, \eg
\beq[mradionfinal]
\De\scr{\scr{L}} \sim \frac{g^2}{16\pi^2} \frac{\hat\om}{\LIR}
F^{\mu\nu} F_{\mu\nu},
\eeq
where $\hat\om$ is the canonically normalized radion field and
$F_{\mu\nu}$ is a standard model field strength.
The radion mass is given by \Eq{mradion} with $n_1 = \tilde{d}$:
\beq
m_{\rm radion} \sim \LIR \om^{\tilde{d} - 4}.
\eeq
For the large values of $\LIR$ we are considering, the radion effectively
decouples in collider experiments.

There are two scenarios for SUSY breaking that we will discuss.
We first consider the case where SUSY is broken in the visible sector
by direct mediation from the CFT.
In the 5D description, the SUSY breaking effects arise from the couplings in
\Eq{directSUSY}.
In this case, we have (see \Eq{MSUSY})
\beq
\om \sim ( 10^{-16} )^{1 / (\tilde{d} - 3)}.
\eeq
For example, for $\tilde{d} = 5$, we have $\om \sim 10^{-8}$, which gives
$\La_{\rm IR} \sim 10^{10}\GeV$.
From \Eqs{mradionfinal} and \eq{MSUSY},
we see that the radion mass is of order $100\GeV$,
for any value of $\tilde{d}$.
There is no model-independent prediction for the pattern of soft masses
in this scenario.
If we make the plausible assumption that the scalar masses and $A$ terms
are universal at the fundamental scale, the SUSY breaking pattern is the
same as in `minimal SUGRA.'

We now briefly consider radion cosmology in this scenario.
We expect radion oscillations to dominate the universe when the
temperature drops below the radion mass of order 100~GeV.
The reheat temperature is of order
\beq
T_{\rm RH} \sim 3\TeV \left( \frac{\LIR}{10^{10}\GeV} \right)^{-1}.
\eeq
This is easily large enough for a realistic cosmology. 

The other scenario we discuss is that SUSY breaking is dominated by a
nonzero constant superpotential.
In this case, $m_{\rm radion} \gg 100\GeV$ and
SUSY breaking in the observable sector is anomaly-mediated.
This also requires additional structure to obtain a realistic
superpartner mass spectrum, but the mechanisms
of \eg\ \Refs{PR,NW} can be used to obtain realistic models.
The detailed phenomenology is model-dependent in this case also.
Radion cosmology is easily realistic.
For example, for $\tilde{d} = 5$ and $\om = 10^{-7}$,
we have $\LIR \sim 10^{11}\GeV$, $m_{\rm radion} \sim 10\TeV$,
and $T_{\rm RH} \sim 10^5 \GeV$.

\section{Conclusions}
We have proposed a new paradigm for SUSY and SUSY breaking.
In this approach, fundamental physics is completely non-supersymmetric,
but the theory flows toward a supersymmetric conformal fixed point
at low energies.
SUSY is therefore an accidental approximate symmetry, and
SUSY breaking at low energies is explicit rather than spontaneous.
Remarkably, many of the features required of a realistic SUSY
model follow very generically from the property that the fixed point
is attractive, \ie\ there are no relevant SUSY breaking perturbations
of the fixed point.
SUSY breaking at low energies naturally arises because the
approach to the fixed point is halted
due to a potential generated by irrelevant operators.
This is contrast to the standard paradigm of spontaneous SUSY breaking
which requires carefully chosen SUSY breaking sectors that are generally not
robust against perturbations.
Also, in the present framework
all required SUSY breaking in the visible sector naturally occurs
at the same scale with no small input parameters.
If there is a small parameter due to an approximate broken $U(1)_R$
symmetry, SUSY breaking
can be dominated by anomaly-mediated SUSY breaking.

The detailed phenomenology of visible sector SUSY breaking is
model-dependent, but there are some general consequences of this framework.
Because SUSY is not an exact symmetry in the UV, there is no Goldstino.
Also, the gravitational sector is completely non-supersymmetric.
This means that there is no gravitino, and hence no problems with gravitino
cosmology.
Also, we expect that there are no gravitational moduli, which cause
grave cosmological difficulties in \eg\ string theory.
There is a dilaton modulus in the CFT, but it couples much stronger
than gravity, and does not give rise to cosmological difficulties.
Finally, this framework requires that the standard model is composite at
a scale
\beq
\La_{\rm IR} \lsim 10^{11}\GeV.
\eeq
This is below the unification scale, but one-step gauge coupling unification
can naturally be explained by accelerated unification \cite{accelunif}.

We conclude that this framework is an attractive alternative to the standard
paradigm of spontaneously broken supersymmetry.

\section*{Acknowledgements}
We thank A. Pomarol and R. Sundrum for helpful discussions, and
D.E. Kaplan and R. Sundrum for comments on the manuscript.
M.A.L. thanks the
Aspen Center for Physics for hospitality during part of this work.
This work was supported by NSF grant PHY-0099544.


\newpage
\begin{thebibliography}{99}

\bibitem{HBN} 
S.~Chadha and H.~B.~Nielsen,
``Lorentz invariance as a low-energy phenomenon,''
Nucl.\ Phys.\ B {\bf 217}, 125 (1983).

\bibitem{Seiberg}
N.~Seiberg,
``Electric--magnetic duality in supersymmetric nonAbelian gauge theories,''
Nucl.\ Phys.\ B {\bf 435}, 129 (1995)
[arXiv:hep-th/9411149].

\bibitem{LR}
M.~A.~Luty and R.~Rattazzi,
``Soft supersymmetry breaking in deformed moduli spaces,
conformal  theories and $N = 2$ Yang-Mills theory,''
JHEP {\bf 9911}, 001 (1999)
[arXiv:hep-th/9908085].

\bibitem{DSB}
Y.~Shadmi and Y.~Shirman,
``Dynamical supersymmetry breaking,''
Rev.\ Mod.\ Phys.\  {\bf 72}, 25 (2000)
[arXiv:hep-th/9907225].

\bibitem{AMSB}
L.~Randall and R.~Sundrum,
``Out of this world supersymmetry breaking,''
Nucl.\ Phys.\ B {\bf 557}, 79 (1999)
[arXiv:hep-th/9810155];
G.~F.~Giudice, M.~A.~Luty, H.~Murayama and R.~Rattazzi,
``Gaugino mass without singlets,''
JHEP {\bf 9812}, 027 (1998)
[arXiv:hep-ph/9810442].

\bibitem{PR} 
A.~Pomarol and R.~Rattazzi,
``Sparticle masses from the superconformal anomaly,''
JHEP {\bf 9905}, 013 (1999)
[arXiv:hep-ph/9903448].

\bibitem{NW} 
A.~E.~Nelson and N.~T.~Weiner,
``Extended anomaly mediation and new physics at 10-TeV,''
arXiv:hep-ph/0210288.

\bibitem{SUSYwoSUGRA}
M.~A.~Luty,
``Weak scale supersymmetry without weak scale supergravity,''
Phys.\ Rev.\ Lett.\  {\bf 89}, 141801 (2002)
[arXiv:hep-th/0205077].

\bibitem{almostnoscale}
M.~A.~Luty and N.~Okada,
``Almost no-scale supergravity,''
arXiv:hep-th/0209178.

\bibitem{PG}
T.~Gherghetta and A.~Pomarol,
``The standard model partly supersymmetric,''
Phys.\ Rev.\ D {\bf 67}, 085018 (2003)
[arXiv:hep-ph/0302001].

\bibitem{BKN} 
T.~Banks, D.~B.~Kaplan and A.~E.~Nelson,
``Cosmological implications of dynamical supersymmetry breaking,''
Phys.\ Rev.\ D {\bf 49}, 779 (1994)
[arXiv:hep-ph/9308292].

\bibitem{LS3} 
M.~A.~Luty and R.~Sundrum,
Phys.\ Rev.\ D {\bf 65}, 066004 (2002)
[arXiv:hep-th/0105137];
M.~Luty and R.~Sundrum,
``Anomaly mediated supersymmetry breaking in four dimensions, naturally,''
Phys.\ Rev.\ D {\bf 67}, 045007 (2003)
[arXiv:hep-th/0111231].

\bibitem{accelunif}
N.~Arkani-Hamed, A.~G.~Cohen and H.~Georgi,
``Accelerated unification,''
arXiv:hep-th/0108089.

\bibitem{ADSCFT}
J.~M.~Maldacena,
``The large N limit of superconformal field theories and supergravity,''
Adv.\ Theor.\ Math.\ Phys.\  {\bf 2}, 231 (1998)
[Int.\ J.\ Theor.\ Phys.\  {\bf 38}, 1113 (1999)]
[arXiv:hep-th/9711200];
S.~S.~Gubser, I.~R.~Klebanov and A.~M.~Polyakov,
``Gauge theory correlators from non-critical string theory,''
Phys.\ Lett.\ B {\bf 428}, 105 (1998)
[arXiv:hep-th/9802109];
E.~Witten,
``Anti-de Sitter space and holography,''
Adv.\ Theor.\ Math.\ Phys.\  {\bf 2}, 253 (1998)
[arXiv:hep-th/9802150].

\bibitem{ADSCFTRS}
H.~Verlinde,
``Holography and compactification,''
Nucl.\ Phys.\ B {\bf 580}, 264 (2000)
[arXiv:hep-th/9906182].
J. Maldacena, unpublished remarks;
E. Witten, ITP Santa Barbara conference
`New Dimensions in Field Theory and String Theory,',
{\tt http://www.itp.ucsb.edu/online/susy
c99/discussion};
E.~Verlinde and H.~Verlinde,
``RG-flow, gravity and the cosmological constant,''
JHEP {\bf 0005}, 034 (2000)
[arXiv:hep-th/9912018].

\bibitem{RS2}
L.~Randall and R.~Sundrum,
``A large mass hierarchy from a small extra dimension,''
Phys.\ Rev.\ Lett.\  {\bf 83}, 3370 (1999)
[arXiv:hep-ph/9905221].

\bibitem{APR} 
N.~Arkani-Hamed, M.~Porrati and L.~Randall,
``Holography and phenomenology,''
JHEP {\bf 0108}, 017 (2001)
[arXiv:hep-th/0012148].

\bibitem{RZ} 
R.~Rattazzi and A.~Zaffaroni,
``Comments on the holographic picture of the Randall-Sundrum model,''
JHEP {\bf 0104}, 021 (2001)
[arXiv:hep-th/0012248].

\bibitem{LS2}
M.~A.~Luty and R.~Sundrum,
``Hierarchy stabilization in warped supersymmetry,''
Phys.\ Rev.\ D {\bf 64}, 065012 (2001)
[arXiv:hep-th/0012158].

\bibitem{confSUGRA}
K.~S.~Stelle and P.~C.~West,
``Minimal Auxiliary Fields for Supergravity,''
Phys.\ Lett.\ B {\bf 74}, 330 (1978);
S.~Ferrara and P.~van Nieuwenhuizen,
``The Auxiliary Fields of Supergravity,''
Phys.\ Lett.\ B {\bf 74}, 333 (1978);
E.~Cremmer, S.~Ferrara, L.~Girardello and A.~Van Proeyen,
``Yang-Mills Theories with Local Supersymmetry:
Lagrangian, Transformation Laws and Superhiggs Effect,''
Nucl.\ Phys.\ B {\bf 212}, 413 (1983).


\bibitem{LS1}
M.~A.~Luty and R.~Sundrum,
``Radius stabilization and anomaly-mediated supersymmetry breaking,''
Phys.\ Rev.\ D {\bf 62}, 035008 (2000)
[arXiv:hep-th/9910202].


\bibitem{LLP} 
W.~D.~Linch, M.~A.~Luty and J.~Phillips,
``Five dimensional supergravity in $N = 1$ superspace,''
Phys.\ Rev.\ D {\bf 68}, 025008 (2003)
[arXiv:hep-th/0209060].


\bibitem{LLP2} 
I.~L.~Buchbinder, S.~J.~Gates, H.~S.~Goh, W.~D.~Linch, M.~A.~Luty, S.~P.~Ng and J.~Phillips,
``Supergravity loop contributions to brane world supersymmetry breaking,''
arXiv:hep-th/0305169.


\bibitem{RSS}
R.~Rattazzi, C.~A.~Scrucca and A.~Strumia,
``Brane to brane gravity mediation of supersymmetry breaking,''
arXiv:hep-th/0305184.


\bibitem{GP} 
J.~Garriga and A.~Pomarol,
``A stable hierarchy from Casimir forces and the holographic  interpretation,''
Phys.\ Lett.\ B {\bf 560}, 91 (2003)
[arXiv:hep-th/0212227].

\bibitem{GW} 
W.~D.~Goldberger and M.~B.~Wise,
``Modulus stabilization with bulk fields,''
Phys.\ Rev.\ Lett.\  {\bf 83}, 4922 (1999)
[arXiv:hep-ph/9907447].

\bibitem{racetrack}
See \eg\ M.~Dine and Y.~Shirman,
``Remarks on the racetrack scheme,''
Phys.\ Rev.\ D {\bf 63}, 046005 (2001)
[arXiv:hep-th/9906246].

\bibitem{MP}
D.~Marti and A.~Pomarol,
``Supersymmetric theories with compact extra dimensions in N = 1  superfields,''
Phys.\ Rev.\ D {\bf 64}, 105025 (2001)
[arXiv:hep-th/0106256].


\bibitem{GM} 
G.~F.~Giudice and A.~Masiero,
``A Natural Solution to the $\mu$ Problem in Supergravity Theories,''
Phys.\ Lett.\ B {\bf 206}, 480 (1988).


\bibitem{SUSYflavor}
M.~Dine, R.~G.~Leigh and A.~Kagan,
Phys.\ Rev.\ D {\bf 48}, 4269 (1993)
[arXiv:hep-ph/9304299];
See \eg\ Y.~Nir and N.~Seiberg,
``Should squarks be degenerate?,''
Phys.\ Lett.\ B {\bf 309}, 337 (1993)
[arXiv:hep-ph/9304307];
R.~Barbieri, G.~R.~Dvali and L.~J.~Hall,
``Predictions From a $U(2)$ Flavour Symmetry in Supersymmetric Theories,''
Phys.\ Lett.\ B {\bf 377}, 76 (1996)
[arXiv:hep-ph/9512388].

\end{thebibliography}
\end{document}